\documentclass[slac_one]{revtex4}
\usepackage{graphicx}
\usepackage{fancyhdr}
\def\nuebar{\rm{\bar{\nu_e}}}
\def\nue{\rm{\nu_e}}
\def\s2tw{\rm{sin ^2 \theta_{W}}}
\def\nuchrad{\rm{\langle r_{\bar{\nu}_e}^2\rangle}}
\def\munubar{\rm{\mu_{\bar{\nu}_{e}}}}

\begin{document}

\title{Studies of Neutrino-Electron Scattering at the Kuo-Sheng Reactor Neutrino Laboratory}

\author{M. Deniz~$^{1,2}$ and 
H.T. Wong~$^2$ ~~~~~ (for the TEXONO Collaboration)}
\affiliation{
$^1$ Department of Physics, Middle East Technical
University, Ankara 06531, Turkey.\\
$^2$ Institute of Physics, Academia Sinica, Taipei 11529,
Taiwan.
}

\begin{abstract}
Studies on $\nuebar - e^{-}$ elastic scattering were performed using 
a 200-kg CsI(Tl) scintillating crystal detector array at the Kuo-Sheng 
Nuclear Power Plant in Taiwan. The measured cross section 
of $R_{exp}=[1.00 \pm 0.32 (stat)] \cdot R_{SM}$ is consistent 
with the Standard Model expectation 
and the corresponding weak mixing angle is derived as 
$\s2tw=0.24 \pm 0.05 (stat) $.
The results are consistent with a destructive interference effect 
between neutral and charged-currents in this process. 
Limits on neutrino magnetic moment of 
$\munubar < 2.0 \times 10^{-10} \mu_{B}$ 
at $90 \% $ confidence level and 
on electron antineutrino charge radius of $\nuchrad < (0.12 \pm 2.07) \times 10^{-32} cm^{2}$ were also derived.
\end{abstract}

\maketitle

\thispagestyle{fancy}

\section{NEUTRINO - ELECTRON SCATTERING}

Neutrino-electron scatterings  ($\nue(\nuebar) - e^{-}$) are 
fundamental electroweak processes which play 
important roles
in neutrino oscillation studies and in probing the
electroweak parameters of the
Standard Model(SM) 
and in the studies of neutrino properties 
such as the electromagnetic moments and charge radius\cite{pdg06}.
The differential cross section for $\nuebar-e^{-}$ 
scattering can be written as\cite{pdg06,kayser79}:
\begin{eqnarray}
\frac{d\sigma _{SM}}{dT}(\nuebar e)& = &
\frac{G_{F}^{2}m_{e}}{2\pi } \left[\left(g_{V}-g_{A}\right)
^{2}+\left( g_{V}+g_{A}+2\right) ^{2}\left(1- \frac{T}{E_{\nu
}}\right) ^{2}-(g_{V}-g_{A})(g_{V}+g_{A}+2)\frac{m_{e}T}
{E_{\nu}^{2}}\right] \label{eq_cross}
\end{eqnarray}
where $T$ is the kinetic energy of the recoil electron, $E_{\nu }$
is the incident neutrino energy and $g_{V}$, $g_{A}$ are coupling
constants which can be expressed as $ g_{V}=-\frac{1}{2}+2\sin ^{2}\theta _{W} $ and $ g_{A}=-\frac{1}{2}$.

The total cross section for $\nuebar-e^{-}$ scattering can be written as
\begin{eqnarray}
\sigma _{SM}  = \int_{T}\int_{E_{\nu }}\frac{d\sigma _{SM}}
{dT}\frac{d\phi}{dE_{\nu}}dE_{\nu}dT = \frac{G_{F}^{2}m_{e}}{2\pi}\left\{
\begin{array}{c}
\left(g_{V}-g_{A}\right) ^{2}I_{1}+\left(g_{V}+g_{A}+2\right)^{2}I_{2} \\
-(g_{V}-g_{A})(g_{V}+g_{A}+2)I_{3}
\end{array}\right\}\text{ \ \ }\label{eq_cross2}
\end{eqnarray}
where $I_{1},$ $I_{2},$ $I_{3}$ are integrals of the function of 1,
$\left( 1-T/E_{\nu }\right) ^{2}$ and $m_{e}T/E_{\nu }^{2}$ over the antineutrino
spectrum and the recoil energy of electron,
respectively. In the low energy neutrino studies we must consider
the electron mass dependent term $I_{3}$ in Eq. \ref{eq_cross2}
because of its significant contribution to the cross
section\cite{kayser79}.

The value of weak mixing angle ($\s2tw$)
was measured precisely at high energy (10-100~GeV) at  
the accelerators, and at lower energy with 
Moller scattering and atomic parity violation experiments\cite{pdg06}.
The interactions $\nue(\nuebar) - e^{-}$
have the additional unique features of being sensitive
to the contributions of 
charged current (CC), neutral current (NC) and
their interference (INT).
The cross-sections of $\nue - e^{-}$ have been measured
at accelerators\cite{lampf}. 
For reactor $\nuebar-e^{-}$, 
the existing data are either controversial\cite{reines,vogelengel}
or with large uncertainties\cite{munuexpt}. 
There is much room for improvement and 
this work is an attempt to bridge this gap.

\section{EXPERIMENTAL SET-UP}
An important component of
the TEXONO research program is
to study $\nuebar-e^{-}$ elastic scattering at the 
MeV reactor neutrino range. 
The neutrino laboratory is located at the Kuo-Sheng
Nuclear Power Plant a distance of 28~m 
from the reactor core with 2.9~GW of thermal power,
having a total flux of about $6.4\times 10^{12} ~ cm^{-2}s^{-1}$.
The details of neutrino source and neutrino spectrum were discussed in 
Ref.~\cite{texono11}. 
The CsI(Tl) scintillation detector array is enclosed by $4\pi$ 
low-activity passive shielding materials with a total mass of 50~tons,
as well as a layer of active cosmic-ray (CRV) plastic scintillator panels.
The entire target space is covered by a plastic bag flushed with dry 
nitrogen to suppress background due to the diffusion of the radioactive radon gas\cite{texono12}.

\begin{figure}[hbt]
\begin{minipage}{18pc}
\includegraphics[width=15pc]{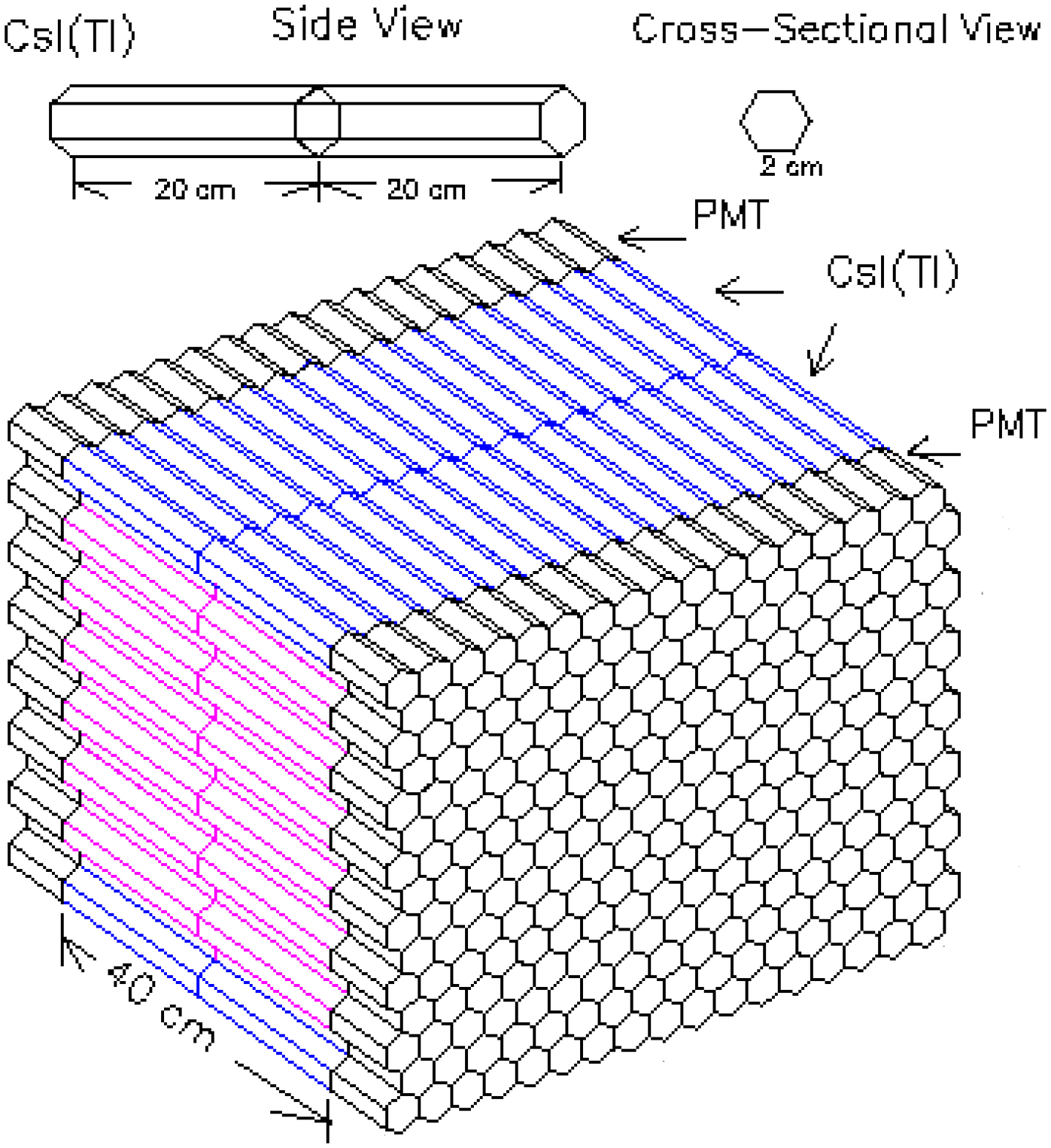}
\caption{\label{csiarray}
Schematic diagram of the CsI(Tl) crystal scintillator array.
}
\end{minipage}\hspace{2pc}%
\begin{minipage}{18pc}
\includegraphics[width=15pc]{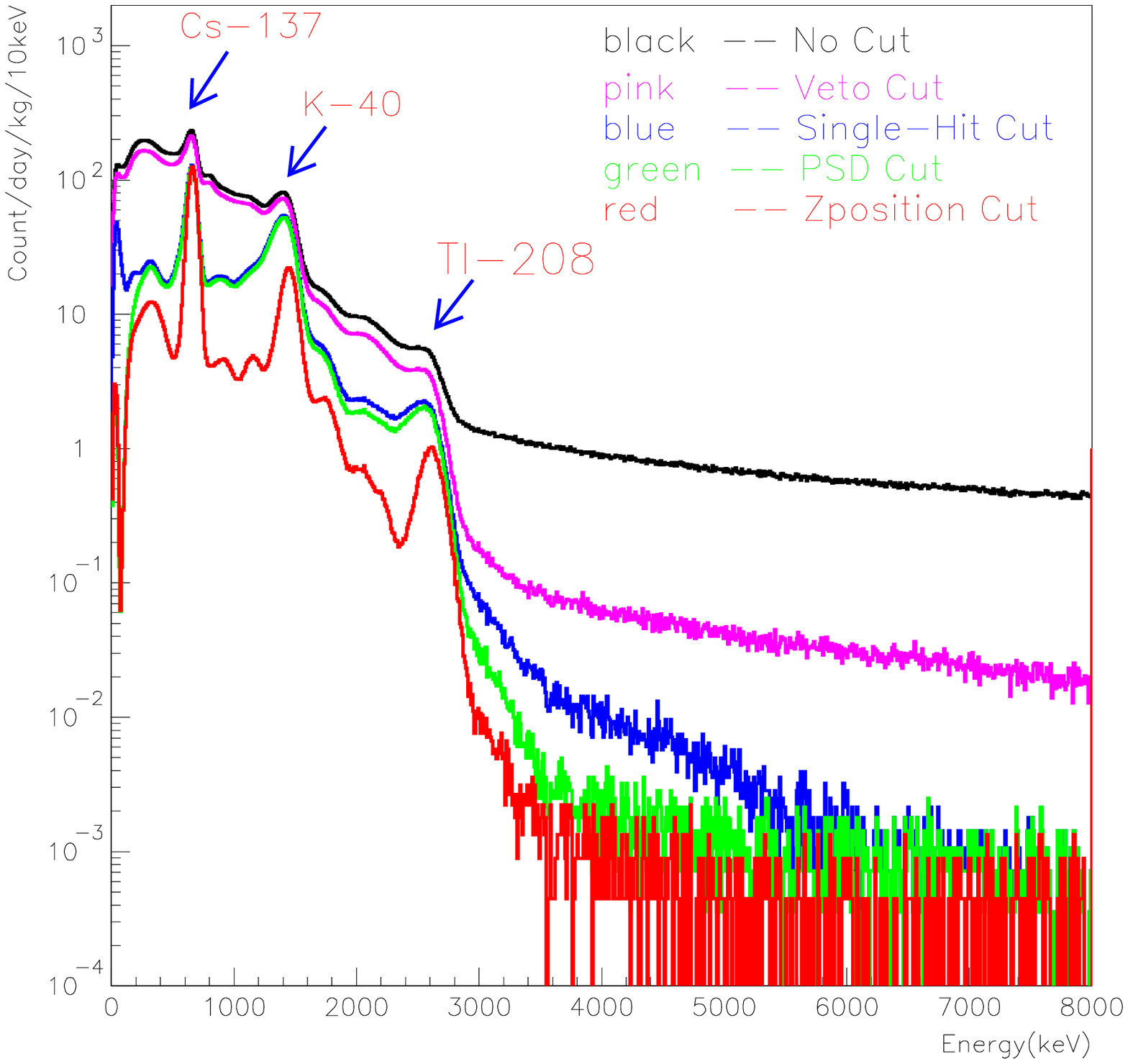}
\caption{\label{spectra}
Measured spectra at the various stages of 
background suppression.
}
\end{minipage}
\end{figure}

The CsI(Tl) crystals were arranged as a $12\times 9$ array matrix 
inside an OFHC copper box, as shown schematically in
Figure~\ref{csiarray}. The detector consisted of 100 crystals
giving a total mass of 200~kg. Each single crystal module has a 
hexagonal-shaped cross-section with 
2~cm side, 40~cm length and 2~kg mass. 
The light output was read out at both ends of the crystal 
by PMTs with low-activity glass of 29~mm diameter.
The properties, advantages and the performance of the prototype modules 
of CsI(Tl) scintillating crystal detector were 
documented elsewhere\cite{texono12,texono13,texono21}. 
These properties make crystal scintillators 
suitable for the study of low energy neutrino experiments. 
The PMT signals were recorded by 
20~MHz Flash Analog-to-Digital-Converters (FADCs)
running on a VME-based data acquisition 
system\cite{texono24}. 
The sum of the two PMT signals gives the
energy of the event, while their difference provides information
on the longitudinal ``Z'' position.
An energy resolution of $< 10\%$ FWHM and a Z-resolution of
$\sim$2~cm at 660~keV as well as excellent 
$\alpha$/$\gamma$ event identification by pulse
shape discrimination (PSD) were demonstrated in 
prototype studies\cite{texono21}.

\section{DATA ANALYSIS}

Neutrino-induced candidate events were selected through
the suppression of: 
(a) cosmic-ray and anti-Compton background
by CRV and multiplicity cuts, 
(b) accidental and $\alpha$- events by PSD, and 
(c) external background by Z-position cut. 
The spectra at the various stages of the background
rejection were displayed in Figure~\ref{spectra}.
In situ calibration was achieved using the  
measured $\gamma$-lines from $^{137}$Cs, $^{40}$K
and $^{208}$Tl.
A signal to background ratio of $\sim$1/15 at 3~MeV was achieved.
The spectra measured during the Reactor OFF periods constituted
a background measurement.

The internal contaminations  of the $^{238}$U and $^{238}$Th series
were measured\cite{yfzhu} and found to be negligible compared to the observed
background rates.
Residual background at the relevant 3$-$6~MeV range are either
cosmic-ray induced or due to coincidence of 
$\gamma$-emissions following $^{208}$Tl decays.
Their intensities were evaluated from the {\it in situ}
multi-hit samples, the $^{208}$Tl-2614~keV lines as well as 
from simulation studies, and the results provide 
the second background measurement.
The background from both methods was
subsequently combined (BKG) and subtracted from the 
candidate Reactor-ON samples.

\section{PHYSICS RESULTS}

A total of 31874.7/7860.1~kg-day
of Reactor ON/OFF
data was recorded and the combined ON$-$BKG 
residual spectrum is displayed in Figure~\ref{residual},
from which various electroweak parameters were derived.
Only events with energy more than 3~MeV above the $^{208}$Tl end-point
were used for physics analysis.
There is an excess in the residual spectrum corresponding
to $\sim$400 neutrino-induced events.
The uncertainties cited in what follows are only statistical.
Intense efforts on the studies of systematic effects are underway.

\subsection{Cross Section and Electroweak Parameters}

\begin{figure}[hbt]
\begin{minipage}{18pc}
\includegraphics[width=15pc]{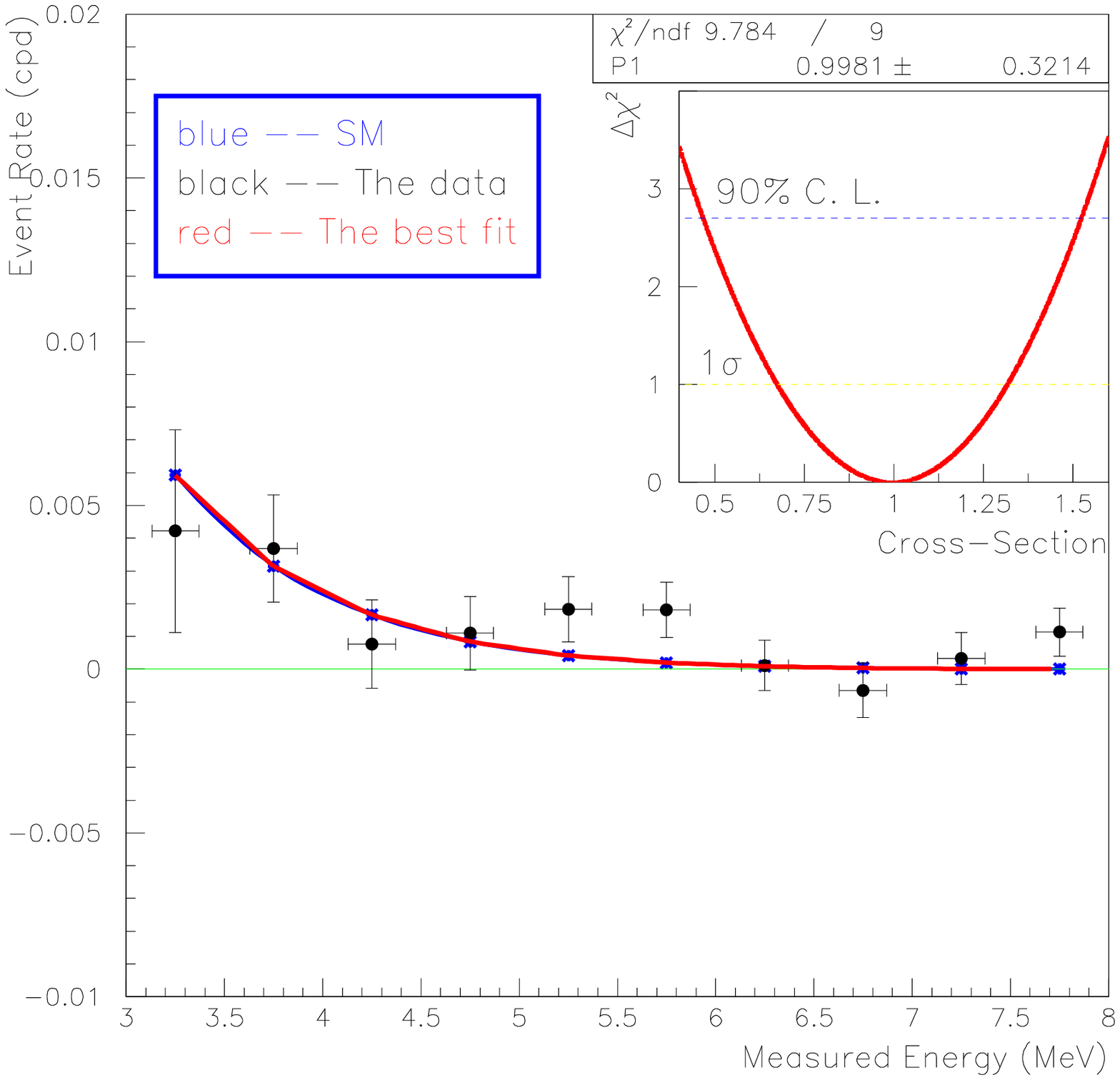}
\caption{\label{residual}
The combined ON-BKG residual spectrum.
The best-fit gives identical curve as
the standard model prediction.
}
\end{minipage}\hspace{2pc}%
\begin{minipage}{18pc}
\includegraphics[width=15pc]{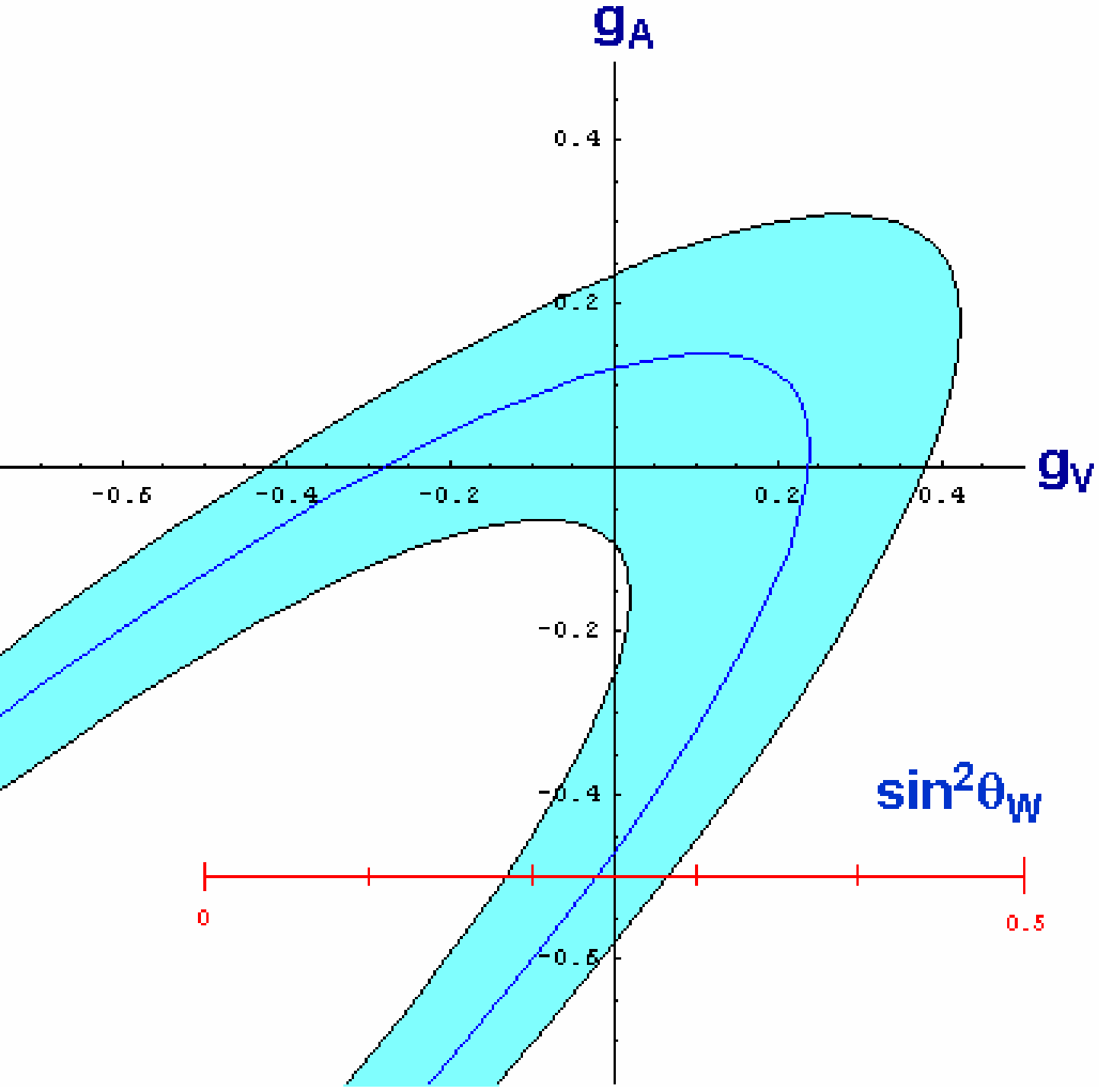}
\caption{\label{gvga}
The 1-$\sigma$ allowed region $ g_{V}-g_{A}$ space, 
together with $\s2tw$. 
}
\end{minipage}
\end{figure}

Denoting the measured event rate as 
\begin{equation}
R _{exp}=\zeta \cdot R_{SM}
\end{equation}
where $R_{SM}$ is the SM predicted values,
the residual spectrum of Figure~\ref{residual} corresponds to 
$ \zeta  = 1.00 \pm 0.32 (stat)  $ with $\chi^{2}/dof = 9.78/9$,
giving 
\begin{equation}
\s2tw = 0.24 \pm 0.05 (stat)  ~~~ . 
\end{equation}
The allowed region in the $g_V - g_A$ plane is depicted in 
Fig. \ref{gvga}.
The accuracy is comparable to that achieved in
accelerator-based
$\nue - e$ scattering experiments\cite{lampf}.

Residual spectra 
from OFF-BKG data 
were extracted  and used for demonstrating
the validity of background understanding and 
the analysis procedures.
The fractional deviation
(OFF-BKG)/OFF = 0.011 $\pm$ 0.018 at $\chi^{2}/dof = 8.23/9 $
indicates excellent agreement with SM expectations
and good systematic control.

\begin{table}[hbt]
\caption{The expected $\zeta$ ratios for the different 
interference scenario and how they are compared
to the measured one.}
\label{tab_inter}
\begin{tabular}{lc}
\\ \hline \hline
Interference &  $\zeta$ \\ \hline
Destructive($\eta=1$) & 1 \\
Constructive($\eta=-1$) & 2.46 \\
No Interference($\eta=0$) & 1.73 \\ \hline
Measurement & $1.00 \pm 0.32 (stat) $
\\ \hline \hline
\end{tabular}
\end{table}

To study the interference term,
the event rate is parametrized as 
\begin{equation}
R_{exp} = R^{CC} + R^{NC} + \eta \cdot R^{INT}
\label{eq_sm_cni}
\end{equation}
where $R^{CC/NC/INT}$ are the SM charged-, neutral currents and
interference contributions, respectively.
Table~\ref{tab_inter} shows the expectations on $\zeta$ for
the possible cases.
The measured value of $\zeta$ verifies the SM
prediction of destructive interference.

\subsection{Magnetic Moment and Neutrino Charge Radius}

Existence of neutrino magnetic moment ($\munubar$) 
would contribute an additional term\cite{vogelengel,munureview}
to the cross-section of Eq.~\ref{eq_cross}:
\begin{equation}
\left( \frac{d\sigma }{dT}\right) _{\mu _{\nu}}=\frac{\pi \alpha
_{em}^{2}\mu _{\nu}^{2}}{m_{e}^{2}}\left[ \frac{1-T_{e}/E_{\nu}}{T_{e}}\right]
\label{eq_mm} ~~.
\end{equation}

Parametrizing the measured event rates as 
\begin{equation}
 R_{exp} = R_{SM} + \kappa^{2} \cdot R(\mu_{\nu} = 10^{-10}~\mu_B) ~~ ,
\end{equation}
the best fit value of $ \kappa^{2} = -0.52 \pm 2.74$ at 
$\chi^{2}/dof = 9.79/9 $ was obtained. 
A limit of 
\begin{equation}
\munubar < 2.0 \times 10^{-10} \times \mu_{B}
\end{equation}
at 90\% CL was derived. 

A finite neutrino charge radius $\nuchrad$ would lead to radiative
corrections\cite{vogelengel,rashba} which modify the
electroweak parameters by:
\begin{equation}
g_{V}  \rightarrow  
-\frac{1}{2}+2\s2tw + (2\sqrt{2}\pi\alpha_{em}/3G_F) \nuchrad 
~~~~~  ; ~~~~~
\s2tw   \rightarrow  \s2tw + (\sqrt{2}\pi\alpha_{em}/3 G_F)\nuchrad 
\label{eq_new_sin2}
\end{equation}
where $\alpha_{em}$ and $G_F$ are the fine structure and Fermi constants,
respectively. Results of
\begin{equation}
\nuchrad = (0.12 \pm 2.07) \times 10^{-32} ~cm^{2}
\label{eq_charge_rad2}
\end{equation}
at $\chi2 / dof = 9.82/9$ were derived accordingly.


\begin{thebibliography}{99} 
\bibitem{pdg06} W. M. Yao et al., \textit{J. Phys.} {\bf G 33}, 321 (2006), for details and references.
\bibitem{kayser79} B. Kayser et al., \textit{Phys. Rev.} \textbf{D 20}, 87 (1979).
\bibitem{lampf}
R. C. Allen et. al., \textit{Phys. Rev.} \textbf{D 47}, 11 (1993);
L. B. Aurbach et. al., \textit{Phys. Rev.} \textbf{D 63}, 112001 (2001).
\bibitem{reines}
F. Reines, H.S. Gurr, and H.W. Sobel, \textit{Phys. Rev. Lett.}, 
\textbf{37}, 315 (1976).
\bibitem{vogelengel} 
P. Vogel and J. Engel, \textit{Phys. Rev}. \textbf{D 39}, 3378 (1989).
\bibitem{munuexpt} 
G. S. Vidyakin et al., \textit{JETP\ Lett.} \textbf{55}, 206 (1992);
A. I. Derbin et al., \textit{JETP\ Lett.} \textbf{57}, 796 (1993);
Z. Daraktchieva et al., \textit{Phys. Lett.} \textbf{B 615}, 081601 (2005);
A.G. Beda et al., \textit{arXiv:0705.4576v1} (2007).
\bibitem{texono11} H. T. Wong et al., \textit{Phys. Rev.}
\textbf{D 75}, 012001 (2007).
\bibitem{texono12} H. B. Li et al., \textit{ Nucl. Instrum. Methods} \textbf{A 459}, 93 (2001).
\bibitem{texono13} H. T. Wong et al., \textit{Astroparticle Phys.} \textbf{14}, 141 (2000).
\bibitem{texono21} Y. Liu et al., \textit{Nucl. Instrum. and Methods} \textbf{A 482}, 125 (2002).
\bibitem{texono24} W. P. Lai et al., \textit{Nucl. Instrum. and Methods} \textbf{A 465}, 550 (2001).
\bibitem{yfzhu}
Y.F.~Zhu et al., 
\textit{Nucl. Instrum. Methods} \textbf{A 557}, 490 (2006)
\bibitem{munureview}
H.T. Wong and H.B. Li, 
\textit{Mod. Phys. Lett.} \textbf{A 20}, 1103 (2005).
\bibitem{rashba} J. Barranco et al., \textit{Phys. Lett.} \textbf{B 662},
431 (2008).
\end{thebibliography}
\end{document}